# Characterizing Long-Range Dependencies in Knee Joint Contact Mechanics: A Comparison of Topology Diffusion, Global Routing, and Hybrid Graph Neural Networks

Zhengye Pan[a,b], Jianwei Zuo[a,b], Jiajia Luo[*] [a,b]

[a] Biomedical Engineering Department, Institute of Advanced Clinical Medicine, Peking University, Beijing, China

[b] Institute of Medical Technology, Peking University Health Science Center, Peking University, Beijing, China

[*] Corresponding author, jiajia.luo@pku.edu.cn

**Abstract**

Finite element analysis of knee joint contact mechanics is computationally expensive, which has motivated the development of graph neural network surrogate models. However, effectively representing long-range dependencies in joint mechanical responses remains challenging. This study systematically compared topology diffusion, global routing, and their hybridization for surrogate modeling of knee joint contact mechanics. Using kinematic and force data from nine soccer players performing change-of-direction maneuvers, finite element simulations were used to generate graph-structured samples for training and evaluation under a grouped three-fold cross-subject evaluation framework. Five architectures were compared: standard MeshGraphNet, hierarchical MeshGraphNet, a routing-only transformer, a topology-biased routing transformer, and a hybrid model. The hybrid model achieved the best overall performance, yielding the lowest full-field error and peak stress error, together with the highest spatial agreement for high-risk regions. Among the non-hybrid models, the standard topology-diffusion model performed best overall, whereas routing-only strategies were less effective. These findings indicate that topology diffusion provides a robust basis for surrogate modeling of knee joint contact mechanics within the present benchmark, while the addition of global routing can further improve reconstruction of clinically relevant high-stress patterns.

# Introduction

The distribution of internal contact stress within the knee joint, together with the formation and migration of high-stress regions, is closely associated with meniscal injury, cartilage degeneration, and the onset and progression of osteoarthritis [1,2]. Finite element analysis (FEA) enables tissue-scale characterization of the multi-tissue coupled contact processes within the joint; however, its complex preprocessing requirements and substantial computational cost limit its applicability in large-scale and near-real-time settings [3,4]. Surrogate models based on message-passing graph neural networks, such as MeshGraphNet (MGN), can exploit node–edge topological relationships to learn the mapping from boundary conditions to full-field mechanical responses. Compared with conventional node-wise regression methods, these models offer clear advantages in predicting joint contact mechanics [5,6] and have shown strong potential for mesh-based physical field prediction [7]. Nevertheless, the contact mechanical response of the knee joint is inherently nonlocal: local stress states are shaped not only by neighboring information, but also by the combined effects of global geometric constraints, distal loading, inter-tissue contact conditions, and load-transfer pathways [8]. Therefore, effective modeling of long-range dependencies is a key factor governing surrogate performance in joint contact mechanics.

To address this issue, existing approaches can generally be grouped into two categories. The first is topology-diffusion modeling, in which information is progressively propagated along the mesh topology through multistep message passing, hierarchical pooling, or coarse-scale interactions, thereby enlarging the receptive field and approximating nonlocal load-transfer processes [9]. The second is global retrieval- or routing-based modeling, in which global tokens or attention mechanisms enable information exchange between distant regions under relatively weak topological constraints [10]. The former emphasizes physical topological priors, whereas the latter emphasizes content-driven global interactions. However, it remains unclear whether the long-range dependencies in joint contact mechanics are better characterized by topology diffusion or by global routing. In addition, it remains unknown whether global routing requires an explicit topological bias and whether diffusion-based and routing-based mechanisms are complementary.

To this end, the present study selected the change-of-direction (CoD) maneuver, a representative task associated with elevated high-stress risk, as the target application scenario [11,12]. A unified five-model comparison framework was established, including the standard MGN, hierarchical MGN (Hi-MGN), Routing-only Token Transformer (RT-MGN), Topology-biased Routing Transformer (TR-MGN), and a hybrid model (Hy-MGN). Under a grouped three-fold evaluation setting with cross-subject separation between training and testing, we systematically compared the performance of topology diffusion, global routing, and their hybridization in surrogate modeling of knee joint contact mechanics, to provide architectural guidance for the design of surrogate models for joint contact mechanics.

## 2 Methods

### 2.1 Dataset construction

#### 2.1.1 Participant recruitment and data acquisition

A total of nine adult male soccer players were recruited in this study (174.4 ± 4.3 cm; 72.43 ± 7.1 kg; 22.1 ± 1.7 years). Eligibility criteria required more than 7 years of soccer training experience. In addition, participants were required to have no history of knee pain or previous knee surgery, no evident lower-limb motor dysfunction, and a dominant right leg. All participants were instructed to refrain from competition within 48 h before testing. Before the experiment, each participant was informed of the study's purpose and procedures, and written informed consent was obtained from all participants. The study protocol was approved by the University Ethics Committee (TY202212088).

Kinematic data were collected using a Vicon three-dimensional motion capture system (200 Hz; V5, Oxford, UK), and ground reaction force (GRF) data were synchronously recorded using a Kistler force platform (1000 Hz; 9286AA, Winterthur, Switzerland). The reflective marker placement protocol followed the setup described by Pan et al. [13]. Before formal testing, each participant completed a 5-min warm-up consisting of jogging (2.5 m/s) and self-directed stretching. During the formal experiment, participants were instructed to perform five 90° CoD trials from the same preparatory posture with an approach speed of 5 ± 0.5 m/s. Of these, the three trials with approach speeds closest to the target value and with full foot contact on the force platform were selected for subsequent analysis. An infrared speed monitor was positioned 5 m from the center of the force platform to measure approach speed, with its height aligned with the participant's greater trochanter. Initial contact was defined as the instant when the vertical ground reaction force (vGRF) exceeded 20 N, and toe-off was defined as the instant when the vGRF fell below 20 N; the interval between these two events was defined as the stance phase.

**2.1.2 Data processing**

The recorded kinematic marker trajectories and GRF data were filtered using a fourth-order zero-lag Butterworth low-pass filter with a cutoff frequency of 10 Hz. Based on the Rajagopal musculoskeletal model, a knee joint model with three degrees of freedom—flexion-extension, internal-external rotation, and abduction-adduction—was implemented in the OpenSim (v4.5, Stanford University, USA) [14], followed by inverse kinematics and joint reaction force analyses. The spatial pose and loading state of the knee joint were both described in a unified tibial anatomical coordinate system, thereby providing standardized boundary conditions for subsequent mechanical simulations.

**2.1.3 Finite element simulation and dataset generation**

In this study, the finite element model was constructed using the standardized topological mesh provided by OpenKnee(s) [3]. Quasi-static simulations were performed in FEBio (2.10, University of Utah, USA) with a time step of 0.01. Specifically, cartilage, ligaments, and menisci were assigned isotropic or transversely isotropic Mooney–Rivlin hyperelastic material properties as appropriate, and ligament pretension was taken into account. For the boundary conditions, knee joint kinematics computed in OpenSim were prescribed as kinematic constraints. In contrast, joint reaction forces were applied as equivalent loads to a reference node located at the distal femur.

Each time step of the finite element simulation was encapsulated as a graph-structured sample for training the surrogate model. For each finite element time step, the graph sample was defined as $G = (V, E)$, where each graph node $v_i \in V$ represented the centroid of a finite element and each edge $e_{ij} \in E$ represented adjacency in the mesh topology. Accordingly, the graph was element-centered rather than based on the original finite element mesh nodes. Node features consisted of the initial geometric position of each element together with the global conditioning variables at the corresponding time step. The global conditioning variables, including joint kinematic and loading variables, were broadcast to all graph nodes. Edge features were defined as the relative position vectors between the centroids of adjacent elements. The prediction target was the von Mises stress associated with each element-centered graph node. Before training, the output stress values were processed using a logarithmic transformation followed by Z-score normalization to improve numerical balance and optimization stability.

## 2.2 Deep surrogate models

### 2.2.1 Network architectures

All five models constructed in this study adopted the same input-output definition and the overall encode-process-decode paradigm, differing only in how long-range dependencies were modeled within the core processor. Specifically, raw node features were first mapped into latent representations through a shared node encoder. Different types of processors then performed information propagation and long-range dependency modeling, and element-wise stress predictions were finally generated through a unified decoder (Fig. 1). As the topology-diffusion baseline, MGN employed a standard message-passing mechanism to propagate information progressively over the mesh topology [7]. At each layer, the model first updated edge representations based on the states of connected nodes and the associated edge features, and then aggregated edge messages to update node states. Through iterative message passing over multiple steps, the receptive field of each node representation gradually expanded along the finite element mesh topology, thereby enabling approximate modeling of nonlocal mechanical transmission. This model represents the basic paradigm in which long-range dependencies are captured through stepwise diffusion over the topology.

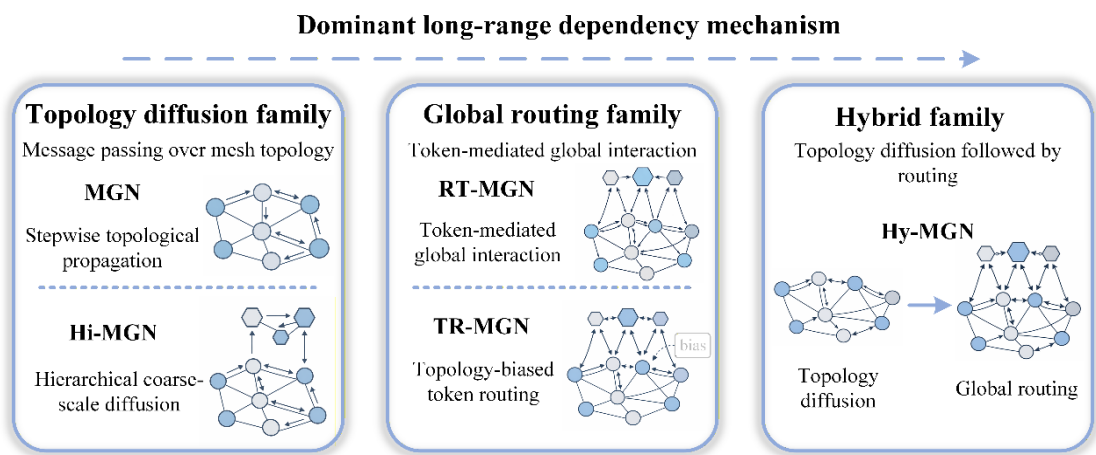

**Figure 1.** Processor taxonomy of the five compared architectures. The models were organized into three families based on their dominant long-range dependency modeling mechanism: topology diffusion, global routing, and hybrid. MGN and Hi-MGN rely on explicit propagation over a mesh topology; RT-MGN and TR-MGN use token-mediated global routing; and Hy-MGN combines topology diffusion with subsequent global routing. Circles denote element-centered graph nodes, solid lines denote mesh-topology connectivity, polygons denote global tokens, and arrows indicate information flow.

Hi-MGN introduced a hierarchical coarse-scale interaction mechanism on top of standard message passing. Specifically, the model constructed a coarse-scale graph by clustering fine-scale graph nodes, performed additional message passing in the coarse-scale space to model interactions between distant regions, and then projected the coarse-scale information back to the original fine-scale node representations. By combining fine-scale local propagation with coarse-scale global propagation, this architecture enhanced the modeling of long-range dependencies. It represented a structured long-range modeling strategy within the topology-diffusion paradigm.

RT-MGN replaced explicit topology-based propagation with a token-based global retrieval-and-routing mechanism. In addition to node latent representations, the model introduced a small set of learnable global tokens. Global information was first aggregated via cross-attention from nodes to tokens, then interactions between distant regions were modeled via self-attention among tokens, and finally, the global context was written back to the node representations via cross-attention from tokens to nodes. Unlike the previous two models, the core processor of RT-MGN did not rely on stepwise propagation along mesh edges; instead, it modeled long-range dependencies as an information retrieval and routing process mediated by a small number of global units.

TR-MGN further incorporated topological bias into RT-MGN to preserve physical structural priors. Specifically, attention biases were constructed based on the graph-topological distance between nodes and token anchors, so that interactions between topologically proximal regions were assigned higher weights. In contrast, interactions between distant or unreachable regions were suppressed. In this way, TR-MGN retained the capacity for global routing while introducing topological constraints derived from the finite element mesh, thereby representing a global routing strategy guided by physical priors.

Hy-MGN combined topology diffusion and global routing in series. The model first employed an MGN processor to perform local-to-mid-range information diffusion along the mesh topology, thereby obtaining node latent representations while preserving the characteristics of physically grounded topological propagation. A token-routing module was then introduced to enable global interaction and contextual reorganization of node representations, allowing the model to further capture long-range dependencies that might be difficult to represent adequately within a finite number of message-passing steps. This model was used to examine whether topology diffusion and global routing are complementary and whether their combination can further improve predictive performance for contact mechanical fields.

**2.2.2 Training and inference**

This study adopted a grouped three-fold cross-validation design with cross-subject separation between training and testing. The nine participants were divided into three predefined subject groups: Fold 1 (P1-P3), Fold 2 (P4-P6), and Fold 3 ($P_7$-$P_9$). In each fold, one group was held out as the test set, and the remaining two groups were used for model development. Within the development set, a validation subset was further created by randomly sampling a fixed proportion of time-step graph samples from the training samples (validation ratio = 0.1), using the same random seed for reproducibility. Thus, the validation set was sampled at the graph-sample level from the training subjects rather than defined as an additional subject-held-out split. All models took a graph sample from a single finite-element time step as input, and element-wise von Mises stress was used as the supervisory target.

During training, all models were optimized under the same training configuration. The AdamW optimizer was used with an initial learning rate of $1\times10^{-3}$, and the learning rate was adaptively reduced according to the validation loss using the ReduceLROnPlateau strategy. Gradient clipping was also applied to improve optimization stability. Each model was trained for up to 150 epochs, with early stopping based on the validation loss to prevent overfitting. Because the number of nodes and edges varied across graph samples, the batch size was fixed at 1. The loss function was defined as masked mean squared error (masked MSE), such that prediction error was computed only over nodes corresponding to valid tissue regions, thereby avoiding interference from invalid regions during training.

To ensure that performance differences could be attributed to the long-range dependency modeling mechanism itself rather than to differences in input representation or training settings, all five models used the same node and edge feature definitions, a shared encode–decode framework, the same hidden dimensionality, the same data split, loss function, optimizer, and early stopping strategy [15]. For models centered on topology diffusion, MGN, Hi-MGN, and Hy-MGN all used three message-passing steps. For models containing routing modules, RT-MGN, TR-MGN, and Hy-MGN all used three token-routing layers, 16 global tokens, and four attention heads. TR-MGN differed only by the introduction of topological bias to modulate the interaction weights between nodes and tokens. In contrast, Hi-MGN differed only by the addition of a hierarchical coarse-scale propagation module. It should be noted that, because topology diffusion and global routing differ fundamentally in their computational graph structures, this study did not enforce exact equivalence in parameter count or GPU memory usage across paradigms. Instead, a controlled comparison was conducted under a unified core width and consistent training configuration.

**2.2.3 Evaluation metrics**

To comprehensively evaluate the models' ability to reconstruct the stress field, while emphasizing sensitivity to errors arising from long-range dependencies and clinically relevant high-stress regions, comparisons were conducted from three complementary perspectives: full-field error, peak/high-quantile fidelity, and spatial consistency. For full-field error, root mean square error (RMSE) and mean absolute error (MAE) were used to quantify the overall prediction error, and normalized RMSE (nRMSE) was further introduced to reduce the influence of physical scale. Let $N$ denote the total number of graph nodes at time step $t$. Because each graph node corresponded to the centroid of a finite element, $y_i$ and $\hat{y}_i$ denote the ground-truth and predicted element-wise von Mises stress, respectively, for graph node $i$. The metrics were calculated as follows:

$$\text{RMSE} = \sqrt{\frac{1}{N}\sum_{i=1}^{N}(\hat{y}_i - y_i)^2} \tag{1}$$

$$\text{MAE} = \frac{1}{N}\sum_{i=1}^{N}|\hat{y}_i - y_i| \tag{2}$$

$$\mathrm{nRMSE} = \frac{\mathrm{RMSE}}{\max_j(y_j)} \tag{3}$$

Here, nRMSE was normalized by the peak ground-truth stress at the current time step, max(*y*), thereby directly reflecting the proportion of prediction error relative to the instantaneous maximum load level. To specifically assess stress-concentrated regions, the relative error of the peak stress (RE$_{max}$) and the relative error of the 95th-percentile stress (RE$_{P95}$) were defined to examine whether the models exhibited the "peak-shaving" effect commonly observed in deep surrogate modeling:

$$RE_{max} = \frac{|max_i(\hat{y}_i) - max_i(y_i)|}{max_i(y_i)} \tag{4}$$

$$RE_{P95} = \frac{|P_{95}(\hat{y}) - P_{95}(y)|}{P_{95}(y)} \tag{5}$$

$P_{95}(\cdot)$ denotes the 95th percentile of the data distribution. These two metrics are intended to capture prediction errors in the joint's most mechanically vulnerable, high-stress regions. In addition, to evaluate whether the models could correctly localize high-stress risk regions in space, the node-level Pearson correlation coefficient (*r*) was calculated, and the Dice coefficient and intersection over union (IoU) were further introduced to assess the spatial overlap of high-risk regions:

$$r = \frac{\sum(\hat{y}_i - \bar{\hat{y}})(y_i - \bar{y})}{\sqrt{\sum(\hat{y}_i - \bar{\hat{y}})^2}\sqrt{\sum(y_i - \bar{y})^2}} \tag{6}$$

$$\mathrm{Dice} = \frac{2|S_{true} \cap S_{pred}|}{|S_{true}| + |S_{pred}|} \tag{7}$$

$$\mathrm{IoU} = \frac{|S_{pred} \cap S_{true}|}{|S_{pred} \cup S_{true}|} \tag{8}$$

Here, the set $S = \{i \mid y_i > P_{95}(y)\}$ defines the set of high-risk region nodes whose stress values exceed the 95th percentile. Furthermore, to quantify the ability of the models to localize hotspot positions, a hotspot distance metric, $d_t^{\mathrm{hot}}$, was introduced to measure the spatial offset between the ground-truth and predicted high-risk regions. Let $\mathbf{p}_i$ denote the centroid coordinates of graph node *i*, that is, the centroid of finite element *i*. The geometric centers of the ground-truth and predicted high-risk regions were then defined as follows:

$$\mathbf{c}_t^{true} = \frac{1}{|S_t^{true}|} \sum_{i \in S_t^{true}} \mathbf{p}_i \tag{9}$$

$$\mathbf{c}_t^{pred} = \frac{1}{|S_t^{pred}|} \sum_{i \in S_t^{pred}} \mathbf{p}_i \tag{10}$$

and

$$d_t^{hot} = \| \mathbf{c}_t^{true} - \mathbf{c}_t^{pred} \|_2 \tag{11}$$

where $d_t^{\text{hot}}$ denotes the Euclidean distance between the geometric centers of the ground-truth and predicted hotspot regions, and was used to measure their spatial displacement.

**2.3 Statistical analysis**

To evaluate performance differences among the models across multiple metrics, a nonparametric statistical testing framework was adopted. For each metric, values were first aggregated within each subject across all eligible test samples, and the resulting subject-level values were used as the statistical units for hypothesis testing. The Friedman test was then used to examine overall differences among models in terms of global error, peak fidelity, and spatial consistency. When a significant overall effect was detected, post hoc pairwise comparisons were performed using the Wilcoxon signed-rank test, with multiple-comparison correction according to the Holm-Bonferroni procedure. A two-sided significance level of α = 0.05 was used throughout. All metrics are reported as mean ± standard deviation (mean ± SD). Statistical analyses were performed in Python 3.9.13.

# 3 Results

**3.1 Full-field stress reconstruction accuracy**

The prediction errors for full-field stress are summarized in Table 1. Among all models, Hy-MGN achieved the best performance in both RMSE and MAE and significantly outperformed the other four models ($p < 0.05$). Among the remaining models, MGN generally showed the best overall control of full-field error, whereas RT-MGN, Hi-MGN, and TR-MGN exhibited relatively higher errors. Notably, TR-MGN performed the worst on both metrics.

**Table 1.** Comparison of full-field stress prediction errors among the five models (Mean ± SD)

|      | MGN | Hi-MGN | RT-MGN | TR-MGN | Hy-MGN |
|------|-----|--------|--------|--------|--------|
| RMSE | 0.055 ± 0.008 | 0.058 ± 0.006 | 0.059 ± 0.007 | 0.061 ± 0.006 | **0.044 ± 0.005** |
| MAE  | 0.036 ± 0.006 | 0.037 ± 0.005 | 0.036 ± 0.005 | 0.038 ± 0.005 | **0.029 ± 0.004** |

Note: **Boldface** indicates results that are significantly different from all other models within each metric.

Figure 2 shows the normalized reconstruction error and spatial consistency of the five models for full-field stress reconstruction. Hy-MGN achieved both the lowest normalized error and the highest consistency, and significantly outperformed all other models on both metrics ($p < 0.05$). Among the remaining four models, MGN showed the best overall performance, whereas TR-MGN performed the worst on both metrics. Hi-MGN and RT-MGN showed similar Pearson correlation values, with no significant difference between them; however, for nRMSE, Hi-MGN significantly outperformed RT-MGN ($p < 0.05$).

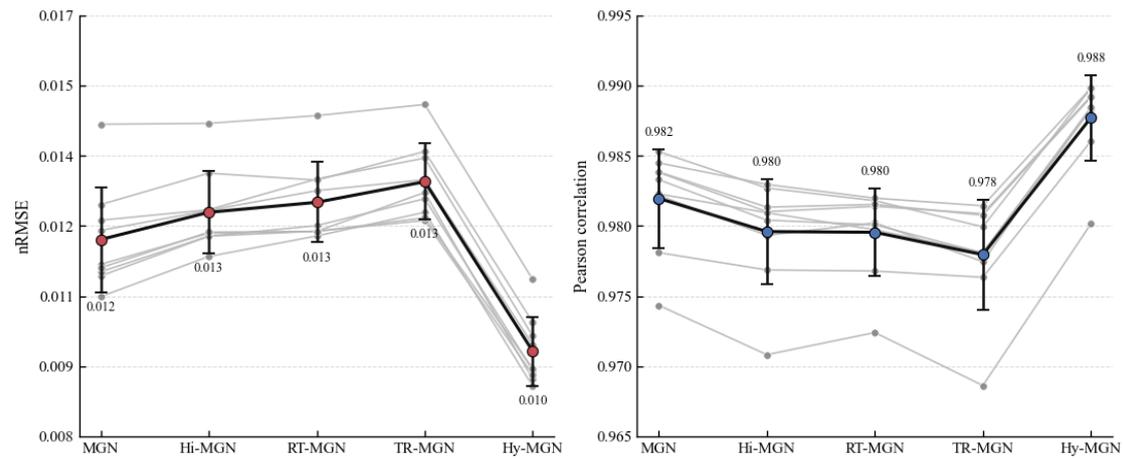

**Figure 2.** Normalized reconstruction error and spatial consistency of the five models for full-field stress reconstruction. Gray lines connect the results from the same subject across models, and gray dots denote individual subject-level values. Colored circles and black connecting lines indicate the group means, and error bars denote the standard deviation.

## 3.2 Accuracy of peak and high-quantile stress reconstruction and their phase-resolved patterns

Figure 3 compares the joint performance of the five models in terms of peak stress relative error ($RE_{max}$) and high-quantile stress relative error ($RE_{P95}$). Overall, Hy-MGN was located in the lower-left region of the plot, indicating simultaneously lower $RE_{max}$ and $RE_{P95}$ and thus the best overall performance. MGN ranked second, whereas TR-MGN occupied the rightmost region and performed the worst overall. For $RE_{max}$, the overall difference among the five models was significant, and post hoc pairwise comparisons further showed significant differences between every pair of models ($p < 0.05$), indicating that different long-range dependency modeling mechanisms had a substantial impact on peak stress reconstruction. In contrast, differences among models along the $RE_{P95}$ dimension were smaller. Although Hy-MGN and Hi-MGN showed relatively lower mean $RE_{P95}$ values, whereas MGN and TR-MGN exhibited greater dispersion, neither the overall comparison nor the post hoc pairwise comparisons reached statistical significance.

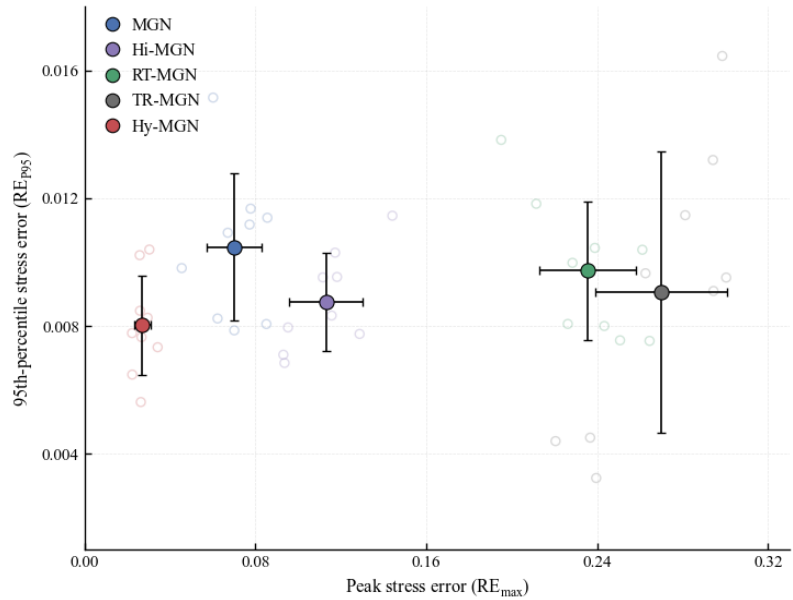

**Figure 3.** Joint comparison of peak stress relative error and high-quantile stress relative error across the five models. Open circles denote individual subject-level results, whereas filled circles and error bars indicate the group mean ± standard deviation. Points closer to the lower-left region indicate better overall performance across the two metrics.

Figure 4 illustrates the phase-resolved patterns of high-quantile error ($RE_{P95}$) and peak error ($RE_{max}$) across the stance phase for the five models. For $RE_{P95}$, Hy-MGN generally maintained lower error throughout the stance phase. RT-MGN and TR-MGN showed similar error levels during the mid-to-late stance, whereas MGN exhibited a relatively higher error region during mid-stance. Overall, differences among models in $RE_{P95}$ were concentrated mainly in the early stance and gradually diminished during the mid-to-late stance.

In contrast, the separation among models was more pronounced for $RE_{max}$. Hy-MGN consistently maintained the lowest peak error throughout the stance phase and showed the most stable distribution. Hi-MGN exhibited overall higher errors than MGN, whereas RT-MGN and TR-MGN persistently showed elevated peak errors across the entire stance phase, with TR-MGN performing worst overall. Notably, compared with $RE_{P95}$, $RE_{max}$ retained clearer model separation during the mid-to-late stance, indicating that peak stress reconstruction was more sensitive to differences in model architecture.

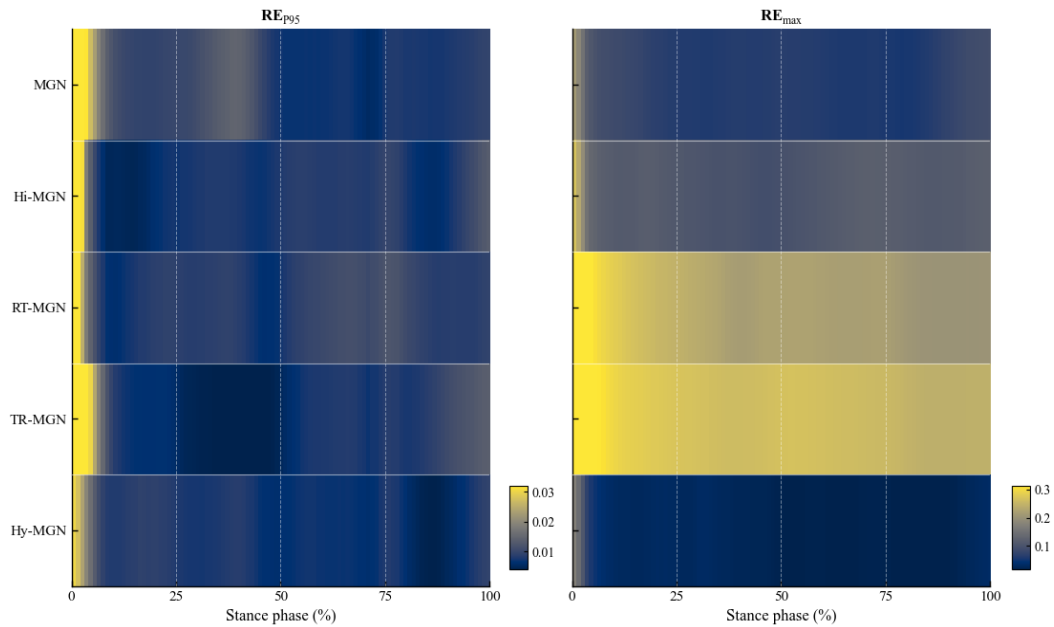

**Figure 4.** Phase-resolved distributions of high-quantile error and peak error across the stance phase for the five models. Darker colors indicate larger errors, and dashed vertical lines mark 25%, 50%, and 75% of the stance phase.

### 3.3 Spatial agreement of high-risk regions

Figure 5 compares the five models in terms of reconstruction agreement for high-risk regions. Hy-MGN achieved the highest agreement on both Dice and IoU and significantly outperformed all other models ($p < 0.05$). In contrast, TR-MGN showed the poorest performance on both metrics. Excluding Hy-MGN, the remaining models showed a monotonic decline in mean performance from MGN to Hi-MGN, RT-MGN, and TR-MGN, with MGN ranking best and TR-MGN worst among the four models.

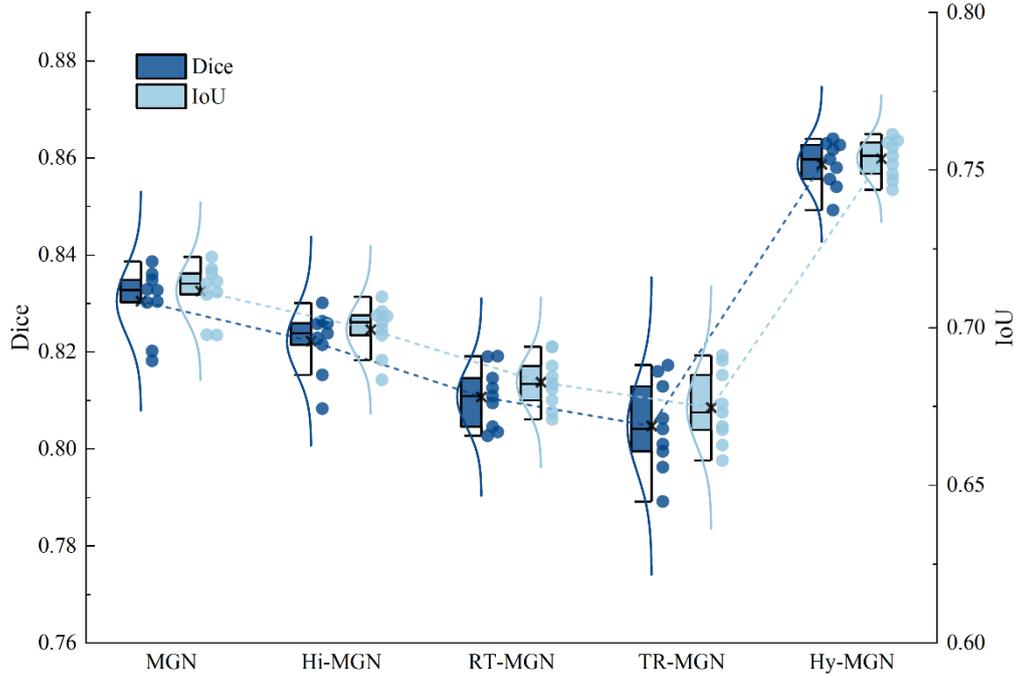

**Figure 5.** Comparison of reconstruction agreement for high-risk regions across the five models. Solid dots represent individual subjects; the overlaid normal distribution curves summarize the across-subject distribution; × denotes the mean; the box indicates the interquartile range, and the line inside the box marks the median.

### 3.4 Hotspot localization ability

The temporal evolution of hotspot localization error during the stance phase for the five models is shown in Figure 6. Hy-MGN maintained a relatively low hotspot localization error throughout most of the stance phase, with comparatively small overall fluctuations. MGN exhibited the most pronounced increase in error during mid-stance and remained above the other models for approximately 40%–70% of the stance phase. Hi-MGN showed an overall temporal pattern similar to that of MGN, but with a lower error magnitude. RT-MGN and TR-MGN displayed comparable error curves during mid-stance and generally fell between MGN/Hi-MGN and Hy-MGN, although TR-MGN still showed noticeable fluctuations in the late stance phase. Toward the end of the stance phase, hotspot localization error decreased further and gradually converged across all models, while Hy-MGN consistently maintained the lowest error.

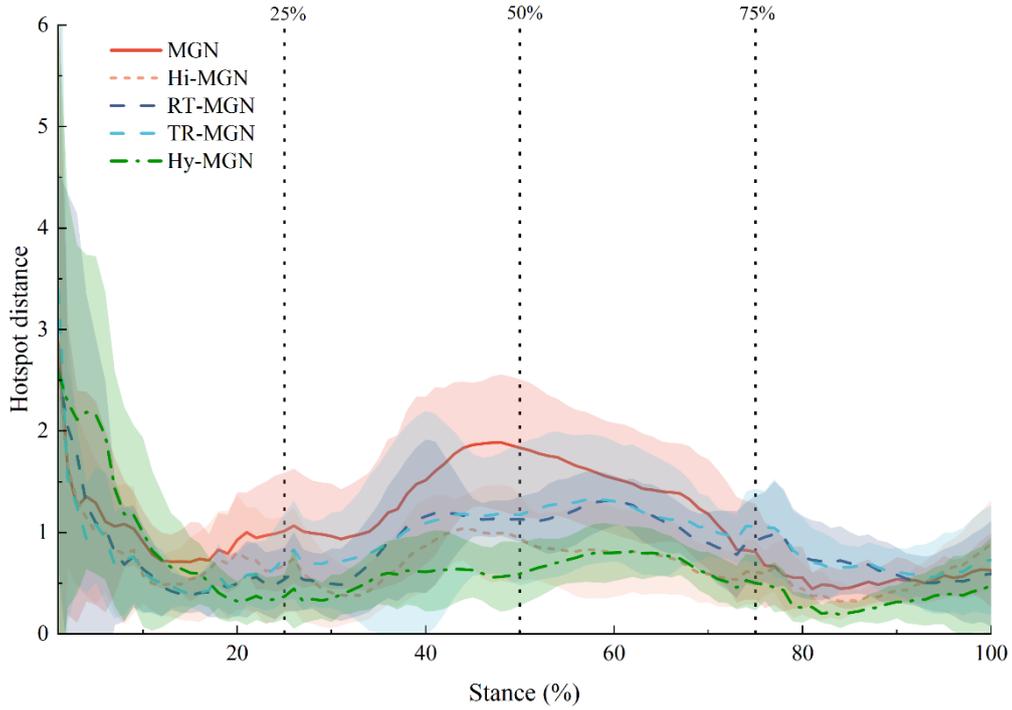

**Figure 6.** Temporal evolution of hotspot localization error during the stance phase (Mean ± SD).

## 4 Discussion

Under a unified experimental framework, this study systematically compared topology diffusion, global routing, and their hybridization for surrogate modeling of knee joint contact mechanics. The results showed that Hy-MGN achieved the best overall performance across the main evaluated metrics, including full-field error, peak stress error, and high-risk region agreement. It also showed a generally favorable temporal pattern in hotspot localization error. Among the non-hybrid models, standard MGN generally ranked best overall. These findings suggest that progressive diffusion over the mesh topology provides a robust basis for modeling long-range dependencies in joint contact mechanics. At the same time, adding global routing can further enhance predictive performance.

The superiority of Hy-MGN indicates that nonlocal dependencies in joint contact mechanics are difficult to characterize adequately with a single mechanism. Topology diffusion progressively propagates information along local connectivity in the finite element mesh. It is therefore more consistent with the physical process by which loads are transmitted and redistributed across multiple tissues within the joint [16]. By contrast, global routing can enhance contextual interaction between distant regions with relatively few layers. It may therefore complement long-range dependencies that cannot be fully captured by a finite number of message-passing steps. Accordingly, the best performance of Hy-MGN suggests that topology diffusion and global routing should not be viewed as mutually exclusive alternatives; rather, they serve different roles in the present task, namely, local, physically grounded propagation and global context integration.

In contrast, the inferior performance of RT-MGN and TR-MGN suggests that global routing alone is insufficient to replace explicit topological propagation. Although token-based mechanisms can establish interactions between distant regions [10], predicting knee joint contact mechanical fields still depends heavily on the mesh topology, because local contact states, tissue continuity, and load-transfer pathways are all structurally constrained by that topology. Without explicit message passing, the models can still achieve global interaction, but they are less able to preserve the structural constraints required for stable local mechanical propagation. This may explain why both RT-MGN and TR-MGN generally underperformed MGN. The fact that TR-MGN did not outperform RT-MGN further indicates that merely modulating attention weights via topological bias is insufficient to replace true stepwise propagation over the topology. In addition, Hi-MGN's inability to surpass standard MGN suggests that a more complex hierarchical structure does not necessarily yield better long-range dependency modeling. In the present task, standard MGN may already provide sufficiently stable coverage of the propagation range most critical for full-field stress reconstruction. In contrast, hierarchical coarse-scale propagation, although expanding the receptive field, may also compromise the fidelity of local geometric and contact information [17], thereby limiting its ability to represent fine-grained stress distributions. This finding suggests that the effectiveness of structural enhancement depends on whether it truly matches the dominant propagation mechanism underlying the target mechanical response, rather than on structural complexity per se.

Several limitations of this study should also be noted. First, the sample size and population diversity were limited, and the task scenario was restricted to a specific movement pattern, which may limit the generalizability of the findings to broader populations and movement repertoires. Second, the present comparison was conducted on a quasi-static finite element surrogate task; under conditions involving stronger temporal dependence or more complex material behavior, the relative advantages of different architectures may change. Finally, although a controlled comparison was performed under unified input definitions, training strategies, and core hidden width, intrinsic differences in parameter count and computational cost remain across paradigms. Therefore, the present results should be interpreted primarily as reflecting the relative performance of different modeling mechanisms in the current task, rather than as a simple comparison between model complexity and predictive performance.

## 5 Conclusion

In surrogate modeling of knee joint contact mechanics, topology diffusion provides a robust foundation for representing nonlocal dependencies, while the addition of global routing can further improve both full-field reconstruction accuracy and the reconstruction of clinically relevant high-stress regions. These findings provide direct architectural guidance for surrogate modeling of joint contact mechanics and suggest that future advances may depend less on choosing between diffusion and routing than on developing more effective ways to integrate the two.

## Funding Statement

This study was supported by the National Key R&D Program of China (grant no. 2023YFC2411201); Beijing Natural Science Foundation (grant no. L259081); NSFC General Program (grant no. 31870942).

## Declaration of Interest Statement

The authors have no conflicts of interest to declare.